\documentstyle[aps,prl,multicol]{revtex}

\begin{document}

\draft

\begin{multicols}{2}
  
\title{Comment on ``Scaling Laws for a System with Long-Range Interactions
       within Tsallis Statistics''}
\author{Benjamin P. Vollmayr-Lee$^{1,2}$ and Erik Luijten$^{2,3}$}
\address{$^{1}$Department of Physics, Bucknell University, Lewisburg, PA 17837,
         USA}
\address{$^{2}$Institut f\"ur Physik, WA 331, Johannes Gutenberg-Universit\"at,
         D-55099 Mainz, Germany}
\address{$^{3}$Max-Planck-Institut f\"ur Polymerforschung, Postfach 3148,
         D-55021 Mainz, Germany}

\date{November 24, 1999}

\maketitle

\pacs{\hspace{-1.8cm}PACS: 05.20.-y, 05.50.+q, 05.70.Ce, 75.10.Hk}

In their recent Letter~\cite{salazar99}, Salazar and Toral (ST) study
numerically a finite Ising chain with non-integrable interactions
\begin{equation}\label{Heq}
  {\cal H}/k_B T = -{1\over 2}\sum_{i\neq j} {Js_is_j\over r_{ij}^{d+\sigma}}
  \;, \qquad (s_i=\pm 1)
\end{equation}
where $-d\leq \sigma\leq 0$~\cite{notation} (like ST, we discuss
general dimensionality~$d$). In particular, they
explore a presumed
connection between non-integrable interactions and Tsallis's non-extensive
statistics.  
 We point out that
   (i) non-integrable interactions provide no more motivation for
   Tsallis statistics than do integrable interactions, i.e., Gibbs statistics
   remain meaningful for the non-integrable case, and in fact provide a
   {\em complete and exact treatment};
and (ii) there are undesirable
features 
of the method ST use to regulate
the non-integrable interactions.

ST study a system of finite length $L$ which is simply
terminated at the ends.  Thus the system energy scales as $E\sim
L^{d+|\sigma|}$ (or $E\sim L^d \ln L$ for $\sigma=0$), and an intensive energy
``density'' is obtained from division by a ``super-extensive volume.''  We show
below that the bulk free-energy density obtained
from this ``non-extensive thermodynamics'' depends explicitly on the regulator,
so all thermodynamics in this boundary-regulated model depend on boundary
effects, and  depend on the system shape as well for $d>1$.  A discussion of 
these issues is lacking in~\cite{salazar99}.

A preferable model employs periodic boundary
conditions, for which a cutoff in the interaction range {\em must\/} be
introduced to regulate the energy.  Then the bulk ``non-extensive
thermodynamics'' depend explicitly on the shape of the cutoff, but not on the
boundaries or the system shape.  

This homogeneous model can be solved exactly, starting from a modified
(\ref{Heq}) with interactions
\begin{equation}
\label{Jeq}
 J \to \biggl\{
       \renewcommand{\arraystretch}{1.2}
       \begin{array}{lc}
       R^{\sigma} w(r_{ij}/R)   & -d\leq \sigma < 0 \\
       (\ln R)^{-1} w(r_{ij}/R) & \sigma = 0 \;,
       \end{array}
\end{equation}
where the cutoff function $w(x)$ decays at least as fast as
$1/x^{|\sigma|+\varepsilon}$ ($\varepsilon>0$) for large $x$, with $w(0)$
finite.  This cutoff enables taking the thermodynamic limit.  The
remaining problem with interactions of range $R$ can be mapped
exactly to a Kac-potential for $\sigma<0$ by identifying
$\phi(x)=w(x)/x^{d+\sigma}$, so that the pair interaction is $R^{-d}\phi(r/R)$
(we present the case $\sigma=0$ elsewhere~\cite{us99}).  One then takes
$R\to\infty$, where the lattice becomes negligible, and recovers the {\em
rigorous\/} result for the free-energy density~\cite{lebowitz66}
\begin{equation}\label{feq}
  f/k_BT = \hbox{convex envelope}\{f^0/k_BT - A m^2\}
\end{equation}
where $f^0$ is the hard-core free energy, $m=L^{-d}\sum s_i$, and
\begin{equation}
\label{Aeq}
  A = \left\{
      \renewcommand{\arraystretch}{1.2}
      \begin{array}{lc}
      {\pi^{d/2}\over\Gamma(d/2)}\int_0^\infty w(x) x^{|\sigma|-1} dx & 
      -d\leq \sigma < 0 \\
      \pi^{d/2} w(0)/ \Gamma(d/2) & \sigma=0
      \end{array}
      \right. \;.
\end{equation}

Next, consider reversing the order of limits, with $R\to\infty$ and
finite $L$.  We present only $d=1$ for clarity.  
The spin $s_i$ interacts with all periodic repeats of $s_j$, leading
to a net pair interaction of
\begin{equation}
\label{eq:eff_pair_potential}
J_{ij} = R^\sigma \sum_{k=-\infty}^\infty
   \frac{w\bigl(\left|kL+r_{ij}\right|/R\bigr)}{|kL+r_{ij}|^{1+\sigma}} \;.
\end{equation}
Remarkably, as $R\to\infty$ this sum converges to the constant value
$(2/L)\int_0^\infty w(x) x^{|\sigma|-1}dx$, thus giving pair interactions that
are independent of spatial separation: the quintessential mean-field theory.
Solving this mean-field theory in the thermodynamic limit gives
exactly the same free energy as before, even for general $d$, thus 
demonstrating that (\ref{feq}) is independent of the order of limits.

Finally, we can treat explicitly the sandwiched case $L \propto R \to \infty$
as well, by use of a hybrid of these two methods, obtaining again
(\ref{feq})~\cite{us99}.  This last limit corresponds directly to
``non-extensive thermodynamics,'' since the $R^{\sigma}$ factor in (\ref{Jeq})
may be interpreted instead as the $L$-dependent temperature employed
in~\cite{salazar99}.  Thus we have solved, with standard methods, the
homogeneous version of ``non-extensive thermodynamics,'' while the
inhomogeneous version studied in \cite{salazar99} would result only in
{\em boundary-dependent\/} modifications to~(\ref{feq}), due to the
system-shape dependence of the cutoff function.

In summary, we find non-integrable interactions as amenable
to Gibbs statistics as integrable interactions, leaving the 
application of alternative methods still with burden of motivation.

\end{multicols}


\begin{references}
\bibitem{notation} We do not follow \cite{salazar99} in using ``long-range'' to
  mean non-integrable, since ``long-range interactions'' already has a standard
  and important meaning: $0<\sigma < 2-\eta_{\rm sr}$.
\end{references}
\end{document}